\documentclass{emulateapj}
\usepackage{apjfonts}
\usepackage{lscape}

\newcommand{\solarmass}{\mbox{${\rm M_{\odot}}$}}

\def\ltsima{$\; \buildrel < \over \sim \;$}
\def\simlt{\lower.5ex\hbox{\ltsima}}
\def\gtsima{$\; \buildrel > \over \sim \;$}
\def\simgt{\lower.5ex\hbox{\gtsima}}

\newcommand{\gro}{{GRO J1655--40}}
\newcommand{\grs}{{GRS 1915+105}}

\newcommand{\etal}{{et al.}}

\newcommand{\asca}{{\it ASCA}}
\newcommand{\rxte}{{\it RXTE}}

\newcommand{\chandra}{{\it Chandra}}

\newcommand{\swift}{{\it Swift}}

\newcommand{\suzaku}{{\it Suzaku}}

\lefthead{Ueda et al.}
\righthead{\grs\ in soft state}

\begin{document}

\title{GRS 1915+105 in ``Soft State'': Nature of Accretion Disk Wind
and Origin of X-Ray Emission}

\author{Yoshihiro Ueda\altaffilmark{1}, 
Kazutaka Yamaoka\altaffilmark{2}, 
Ronald Remillard\altaffilmark{3}}

\altaffiltext{1}{Department of Astronomy, Kyoto University, Kyoto 606-8502, Japan}
\altaffiltext{2}{Department of Physics and Mathematics, Aoyama Gakuin University, Sagamihara, Kanagawa 229-8558, Japan}
\altaffiltext{3}{Department of Physics, Massachusetts Institute of Technology, 
  Cambridge, MA 02138}

\begin{abstract}

We present the results from simultaneous \chandra\ HETGS and \rxte\
observations of the microquasar \grs\ in its quasi-stable ``soft
state'' (or State A) performed on 2007 August 14, several days after
the state transition from ``hard state'' (State C). The X-ray flux
increased with spectral hardening around the middle of the \chandra\
observation, after which the 67 Hz quasi periodic oscillation (QPO)
became significant. The HETGS spectra reveal at least 32 narrow
absorption lines from highly ionized ions including Ne, Mg, Si, S, Ar,
Ca, Cr, Mn, Fe, whose features are the deepest among those ever
observed with \chandra\ from this source. By fitting to the absorption
line profiles by Voigt functions, we find that the absorber has outflow
velocities of $\approx$150 and $\approx$500 km s$^{-1}$ with a
line-of-sight velocity dispersion of $\approx$70 and $\approx$200 km
s$^{-1}$ for the Si XIV and Fe XXVI ions, respectively. The larger
velocity and its dispersion in heavier ions indicate that the wind has
a non-uniform dynamical structure along the line-of-sight. The
location of the absorber is estimated at $\sim (1-3)\times10^{5} r_{\rm g}$
($r_{\rm g}$ is the gravitational radius) from the source, consistent
with thermally and/or radiation driven winds. By taking into account
narrow spectral features detected with \chandra, the continuum spectra
obtained with \rxte\ in the 3--25 keV band can be well described with
a thermal Comptonization with an electron temperature of $\approx 4$
keV and an optical depth of $\approx 5$ from seed photons from the
standard disk extending down to $(4-7) r_{\rm g}$. In this
interpretation, most of the radiation energy is produced in the
Comptonization corona, which completely covers the inner part of the
disk.  A broad ($1\sigma$ width of $\approx 0.2$ keV) iron-K emission
line and a smeared edge feature are detected, which can be explained
by reflection from the accretion disk at radii larger than $400 r_{\rm
g}$. Our data do not require the presence of an extremely blurred
iron-K line or disk emission originating from radii smaller than
$\approx 4 r_{\rm g}$ in this state.

\end{abstract}


\section{Introduction}

Microquasars are key objects to investigate the physics of accretion
onto a black hole in relation with relativistic jet. \grs\ is the most
important, archetype microquasar, from which the first superluminal
motion in our Galaxy was detected \citep[for review, see][]{fen04}. It
is an X-ray binary consisting of a M-K III star and a black hole of
$14\pm4$ \solarmass with an inclination angle of
66$^\circ$--70$^\circ$ at a distance of $\approx 12$ kpc. The
luminosity of \grs\ reaches close to the Eddington limit, indicating
critical mass accretion takes place. With this reason, \grs\ exhibits
unique properties distinct from canonical black hole binaries
(BHBs). In particular, it sometimes shows dramatic temporal/spectral
variations (``oscillation'') occurring in quasi-regular cycles, most
probably due to thermal instability in the accretion
disk. \citet{bel00} phenomenologically classified the X-ray states
into three, States~A, B, and C. \grs\ spends most of time in State~C
where the X-ray spectrum is relatively hard. The underlying physics of
these complex behaviors and the relation to canonical BHBs is far from
being fully understood, even though it is a key question to establish a
general solution of accretion onto black holes at various mass
accretion rates.

Outflow by a disk wind is one of key ingredients in understanding the
dynamics of an accretion flow in disk systems, including X-ray
binaries and active galactic nuclei. \citet{kot00} discovered
absorption line features of iron-K ions in the X-ray spectra of \grs\
with \asca, revealing the presence of a significant amount of highly
ionized plasma in this system. Similar features were reported in many
other X-ray binaries including both black holes
\citep[e.g.,][]{ued98,mil06} and neutron stars
\citep[e.g.,][]{ued01,boi04}, which are now recognized to be a common
feature in accreting sources over a wide range of luminosity. The
origin of such ionized gas has been established to be a disk wind by
the precise measurement of the outflow velocity with \chandra.
The physical mechanism to produce the disk wind is not established
yet. There are at least 3 possibilities that are not exclusive one
another: thermally-driven wind \citep{beg83}, radiation driven wind
\citep[e.g.,][]{pro00}, and/or magnetically driven wind. Based on a
simple argument from an observed column density and an ionization
parameter, the location of the wind has been estimated at $10^{4-5}
r_{\rm g}$ ($r_{\rm g} \equiv GM/c^2$ is the gravitational radius
where $G$, $M$, $c$ is the gravitational constant, mass of the black
hole, and light velocity, respectively.) from the source, consistent
with thermally and/or radiation driven in all the cases (including
\gro) except for \citet{mil06}, who deduced a much smaller radii from
\chandra\ data of \gro, leading to their conclusion that a
magnetically driven wind is the only solution. This argument is
questioned by \citet{net06}, however, making the discussion still
controversial.

Detailed studies of the disk wind in \grs\ is particularly important
to understand the role and origin of the outflow at high mass
accretion rates. Among the different epochs observed with \asca, the
absorption features were the most prominent (equivalent widths of
30--50 eV for H- and He-like Fe ions) in 1994 and 1995 when the
spectrum is soft and the 1--10 keV flux is relatively faint ($\sim$0.3
Crab). We recognize that in these \asca\ observations the source was
in quasi-stable ``soft state'' (State~A), which is found to be quite
rare case in the long term variability from the \rxte /ASM over $>$10
years. By contrast, the iron-K absorption-line features were much
weaker and even undetectable when the 1--10 keV spectrum was hard,
corresponding to State~C (or ``hard state''). This can be explained by
the increasing photon flux responsible for photo-ionization which
almost fully ionized the ions. \citet{lee02} performed a HETGS
observation of \grs\ in State~C and detected absorption lines from
highly ionized Fe ions. However, the features are much weaker, with
equivalent widths of only ~10 eV, compared with those found in State~A
with \asca. Because of the weakness of these features there still
remain uncertainties in the basic properties of the plasma, such as
the bulk velocity and its dispersion (or kinetic temperature).

In this paper, we report the results of high energy resolution
spectroscopy of \grs\ in ``soft state'' (State~A) performed
simultaneously with \chandra\ HETGS and \rxte\ in 2007 August. The
first purpose is to reveal the origin of the disk wind with the best
spectroscopic data ever obtained for absorption-line diagnostics of
the highly ionized plasma in \grs. Secondly, we study the origin of
the continuum emission in this state by utilizing the ideal
combination of \chandra\ and \rxte . The simultaneous HETGS data
enable us to take into account all the spectral features in the
analysis of the \rxte\ PCA spectra, including
interstellar/circumstellar absorptions with anomalous metal
abundances, narrow absorption lines by highly ionized gas, and a broad
iron-K emission line. \S~2 summarizes the observations and data
reduction. The results and discussion are presented in \S~3 and \S~4,
respectively. We assume the distance toward \grs\ of 12.5 kpc and the
inclination of 70$^\circ$ throughout the paper. Quoted errors for
spectral parameters are 90\% confidence level for a single parameter
unless otherwise mentioned.  Multiwavelengths properties of \grs\
including radio and infrared data over a long period covering our
observation are presented by \citet{ara08}.

\section{Observations and Data Reduction}

\subsection{\chandra\ Data}

The \chandra\ HETGS observation of \grs\ was carried out from 2007
August 14 5:22 (UT) to 19:35 for an exposure of 47.4 ksec as a
target-of-opportunity (ToO) observation (obsID 7485). Figure~1 shows
the long term light curves of \grs\ monitored by the \rxte\ ASM in the
2--12 keV band, its hardness ratio $HR2$ between the 5--12 keV and
3--5 keV bands, and that by \swift\ BAT in the 15--50 keV band. To
catch the source in ``soft state'', we set the trigger condition for
the \chandra\ ToO that the following two criteria holds for 2 or more
successive days: (1) the ASM count rate $< 30$ c/s and (2) $ HR2 <
1.1$. As seen from the figure, \grs\ made a state transition around
2007 August 8, entering into a quasi-stable ``soft state'' where the
15--50 keV flux became below the detection limit of the BAT. The
position of the arrows denote our observation epoch. To make the
effects of pile-up least, we adopt {\it graded} telemetry mode and
readout only one-third of the full frame using a subarray. We did not
use ACIS-S0 and S5, because \grs\ is heavily absorbed and has little
photons below $\sim$ 1 keV.

\begin{figure}
\epsscale{1.1}
\plotone{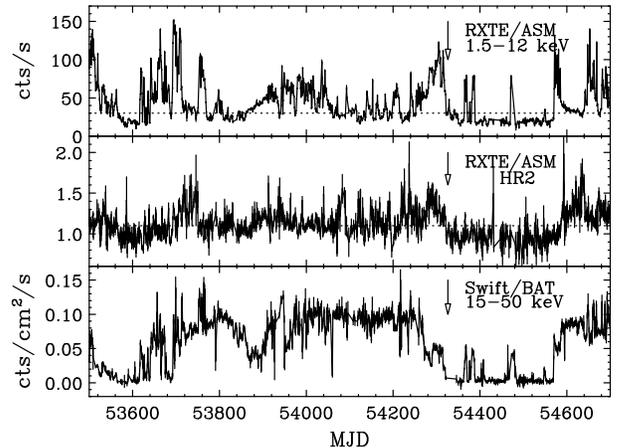}
\caption{
Upper: the long term light curve of \grs\ observed with the \rxte\ ASM. 
Middle: the hardness ratio ($HR2$) of the \rxte\ ASM count rates
between 5--12 keV and 3--5 keV bands. Lower: \swift\ BAT light curve
in the 15--50 keV band.  The epoch of our observations is denoted by
the arrows. The dashed lines correspond to 30 cts/s (upper) and $HR2 =
1.1$ (middle), thresholds for triggering our ToO observations 
aiming at the ``soft'' state.  }
\label{fig1}
\end{figure}

We reduce the data using the CIAO version 3.4 and the CALDB version
3.4.5 provided by the \chandra\ X-ray Center (CXC). To determine the
zero-th order position most accurately under heavy pile-up, we utilize
the {\it tg\_findzo} script\footnote{available from the MIT/CXC home
  page, http://space.mit.edu/cxc/analysis/findzo/findzo.html}.  The
accuracy of relative wavelengths calibration of the HETGS is 0.05 pixel
in position within a chip
\footnote{http://space.mit.edu/CXC/calib/hetgcal.html},
corresponding to 0.00028 $\AA$ in the HEG (or, 14 km s$^{-1}$ at 2 keV
and 47 km s$^{-1}$ at 7 keV), although the accuracy may be worse if we
use the data over multiple chips, because of a chip-gap error
(Marshall, H., private communication). As described below, we are
particularly interested in the energy determination in the 2.0--7.0
keV range (between the $K\alpha$ lines from Si XIV and Fe XXVI), which
is covered within a single chip in the $-1$ order spectrum. We confirm
that the $+1$ data give consistent results with the $-1$ one. Thus,
velocity difference between Si XIV and Fe XXVI ions can be measured
within an accuracy of $<$50 km s$^{-1}$.

In this paper we mainly utilize the 1st-order HEG data, which cover
energies above 4 keV and has twice better energy resolution than the
MEG. For spectra below 1.6 keV, the MEG data are utilized instead, which
have about three times larger effective area there. Pile-up is corrected
according to the method described in \citet{ued04}; we calculate the
fractional count rate loss as $f = 1- {\rm exp} (-7.5 R)$, where $R$ is
the count rate of the ACIS pixel in units of count per pixel per frame,
and multiply the observed spectra by $1/(1-f) \simeq 1 + 7.5 R$. The
correction factor is found to be 15\% at maximum around 4--5 keV in
the HEG data. The spectra from the $+1$ and $-1$ orders are summed. We
conservatively add a systematic error of 3\% in each bin of the
spectra. The XSPEC package (version 11.3.2) is used for spectral
analysis.

\subsection{\rxte\ Data}

Following the trigger of the \chandra\ ToO, {\it Rossi X-ray Timing
Explorer (RXTE)} performed two ToO observations of \grs\ at 2007 August
14 07:36--08:57 (Epoch~I') and 17:00--18:03 (Epoch~II') during the
\chandra\ observation. \rxte\ carries two pointing instruments, the
Proportional Counter Array (PCA: \citealt{jah06}) and the High Energy
Timing Experiment (HEXTE: \citealt{rot98}). We only analyze the PCA
data in this paper, because the spectra of \grs\ were very soft in our
observations and hence the HEXTE data do not give strong
constraints. The HEADAS package (version 6.4) is used for the
analysis. The data are selected by the following criteria: 1) \rxte\
was not in the South Atlantic Anomaly, 2) the elevation angle was
below 10 degrees, and 3) the offset angle between the pointing
direction and the target was less than 0.02 degree. The net exposures
are 2.4 ksec and 3.2 ksec for Epochs~I' and II', respectively.

We utilize the ``standard 2'' PCA data for spectral analysis, which
have 16 sec time resolution and 129 energy channels. We extract energy
spectra from the data of the top layer of Proportional Counter Unit
(PCU) 2, the best calibrated one. The background is subtracted using
the model applicable for bright sources. The deadtime, about 3--4\%,
is corrected. A systematic error of 1.5\% is added for each energy
channel. To check the calibration accuracy, we also analyze the
spectra of Crab Nebula observed on 2007 August 11. We confirm that the
combined PCU~2 and HEXTE (Cluster-B) spectra in the 3.5--200 keV band
are well reproduced by a single power-law with a photon index of
2.07$\pm$0.01 for a fixed Galactic absorption of $N_{\rm H} =
3\times10^{21}$ cm$^{-2}$. The flux in the 3--20 keV band determined
by PCU~2 is 2.63$\times$10$^{-8}$ ergs cm$^{-2}$ s$^{-1}$, which is
consistent within 10\% of the nominal value \citep{too74}.

For timing analysis of the PCA, we use two types of single-bit data
covering the energy range of 3.2--5.6 keV and 5.6--14.8 keV, and
event-mode data covering energies above 14.8 keV. They have a time
resolution of 122 $\mu$s (single bit) and 16 $\mu$s (event mode),
respectively. The light curves from these data are co-added to produce
one light curve in the 3.2--37.9 keV band. The power spectral density
(PSD) is then calculated by the {\it powspec} ftool in the XRONOS
package. It is known that the PSD computed from PCA data in the Leahy
normalization is less than 2, a value expected from Poisson
distribution, at high frequencies above 10 Hz due to deadtime effects
\citep{zha95}. We thus subtract the dead-time corrected Poisson noise
level from the observed PSD, following the method by \citet{mor97}.

\section{Results}

\subsection{Light Curves and Power Spectra}

Figure~2 shows the light curves of \grs\ with 16 sec resolution
obtained with the \chandra\ HETGS in the 1--8 keV band (upper panel),
\rxte\ PCA in the 3.2--38 keV band (middle), and the hardness ratio
between the 5.6--38 keV and 3.2--5.6 keV PCA count rates (lower). In
the upper panel, we plot the sum of the MEG and HEG count rates
including the 1st, 2nd, and 3rd orders. As noticed from the figure,
the averaged flux level was roughly constant in the first half and
then started to increase from the middle of the observation,
accompanied with spectral hardening. The PCU2 count rate increased by
about $70\%$ in the latter observation. It is also seen that the flux
is highly variable on a shorter time scale than $10^4$ sec. For later
discussion, we divide the epoch into two, Epoch I (before 2007 August
14 13:06:40) and Epoch II (after). We refer the two intervals of the
\rxte\ observations as Epoch I' (former) and II' (latter).  These
observations would be classified as ``$\phi$'' (steady State~A) in the
nomenclature of \citet{bel00}, since the hardness ratio values are
low, there is limited variability, and there is no structure in the
color-color diagram (not shown). During Epoch~II, the count rate
approaches the level that is more typical of the ``$\delta$'' class,
where there are transitions between States~A and B, and hence Epoch~II
can be considered to be an interval of ``bright $\phi$'' conditions.

\begin{figure}
\epsscale{1.0}
\includegraphics[angle=270,scale=0.35]{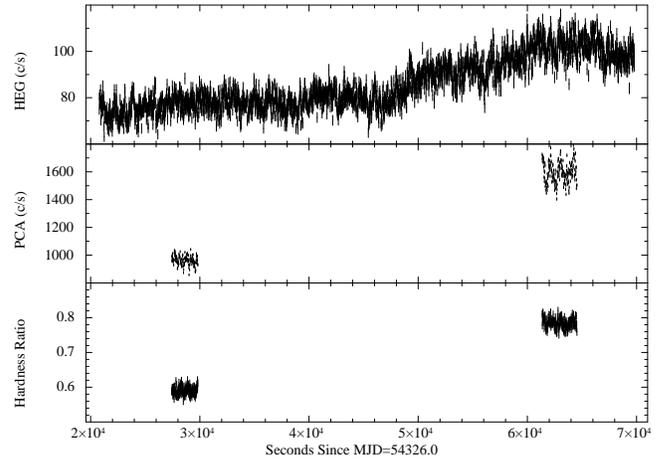}
\caption{
Upper: the \chandra\ HETGS light curve (sum of the HEG 1st, 2nd, and 3rd order events) with a time resolution of 16 sec. No correction is made for pile-up.
Middle: \rxte\ PCA light curve in the 3.2--38 keV band with a time resolution 
of 16 sec (only PCU2 data are utilized).  
Lower: the PCA hardness ratio between the 5.6--38 keV and 3.2--5.6 keV bands.
\label{fig2}}
\end{figure}

The power spectral density in the 3--38 keV band calculated from the
\rxte\ PCA is shown in Figure~3. While there are no clear QPO
features in Epoch I', a QPO is detected at 66.8$\pm$1.1 Hz at
4.3$\sigma$ level in Epoch II'. The coherence parameter in this QPO is
$Q = \nu_0 / {\rm FWHM} \approx 11$, where $\nu_0$ is the central QPO
frequency and the fit uses a Lorentzian profile. We note that the
difference in the QPO intensity between the two epochs is not highly
significant, however. The integrated fractional root mean square
around the 67 Hz QPO is found to be 0.8\%$\pm$0.4\% and
1.6\%$\pm$0.2\% in Epoch I' and II', respectively (the errors are
1$\sigma$).

\begin{figure}
\epsscale{1.1}
\plotone{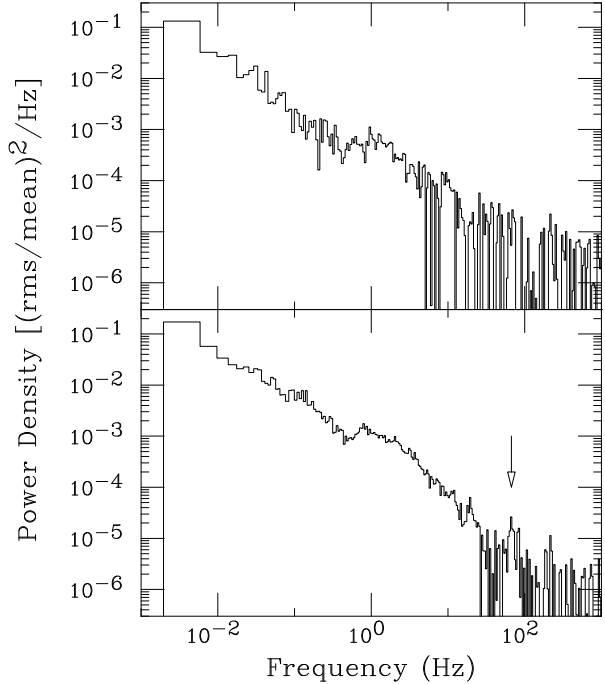}
\caption{
The normalized power spectral density in Epochs I' (upper) and II' (lower)
obtained from the \rxte\ PCA data covering the 3.2--38 keV band. The
Poisson noise level was subtracted after dead time correction. The position
of the 67 Hz QPO is denoted by the arrow.
\label{fig3}}
\end{figure}

\subsection{HETGS Spectra}

Figure~4 shows the HETGS spectra integrated over the whole
observation. Here we only correct the spectrum for effective area as a
function of energy, but not for the instrumental energy resolution,
which is 3.2 eV at 2.0 keV and 40 eV at 7.0 keV in the HEG. The MEG
data are utilized in the 1.0--1.6 keV band, while the HEG data for the
rest. Although the overall spectral shape changed from Epoch I to II
as mentioned above, we first study this spectrum as the base data from
which we can best constrain local spectral features in time-averaged
sense. We find that the overall continuum can be well described by a
cutoff power law model, expressed as $A\;E^{-\Gamma} {\rm exp}(-E/E_{\rm
fold})$, subject to heavy interstellar absorption with non-solar
abundances. The best-fit parameters are summarized in Table~1. Note
that the cutoff power law model is only empirical and does not have
physical meaning itself; we will model the continuum emission by
physical models for the PCA spectra in \S~3.3.

\begin{figure*}
\epsscale{1.1}
\plotone{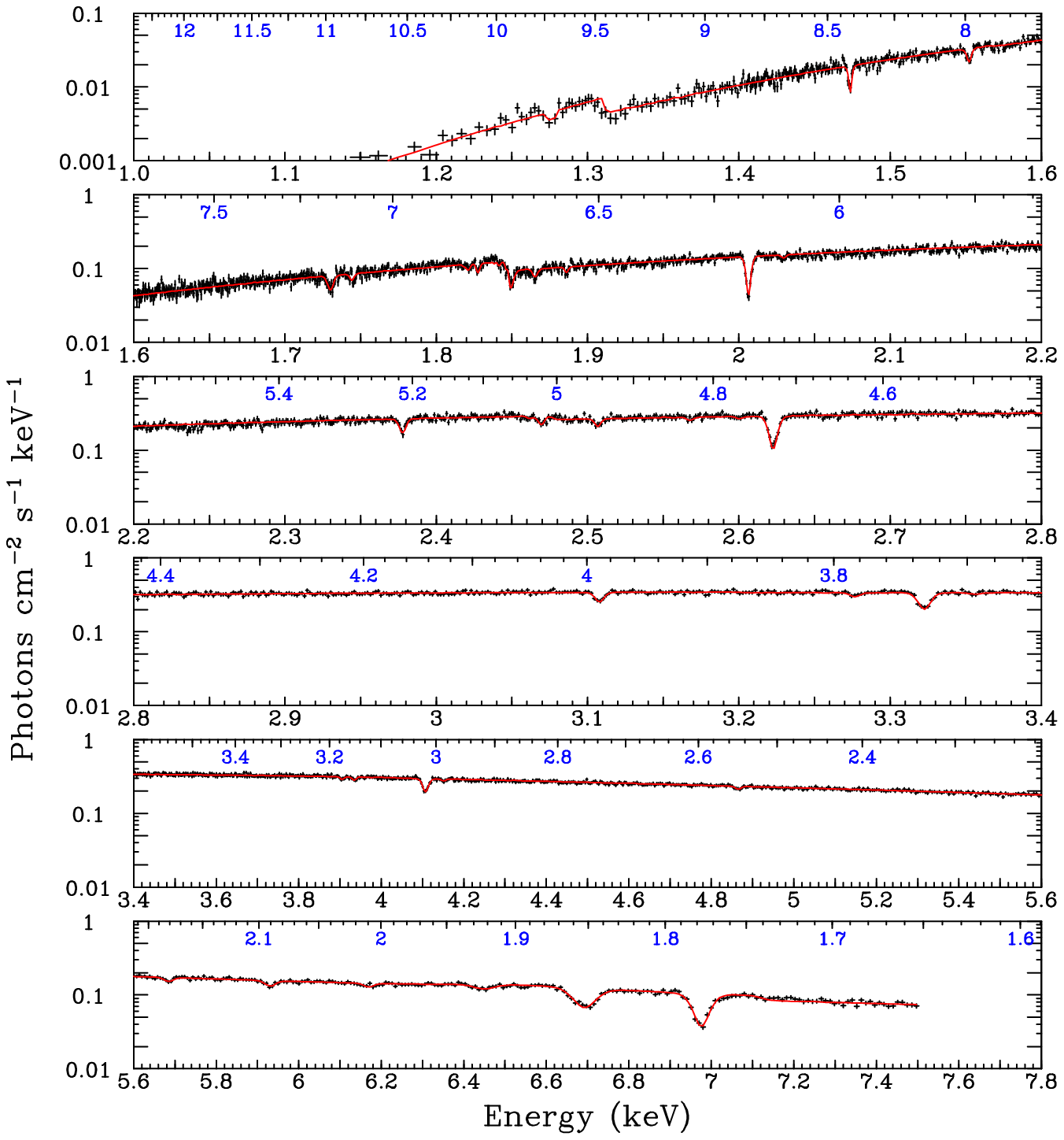}
\caption{The \chandra\ HETGS spectrum integrated over the whole
observation (corrected for the effective area but not for the energy
resolution). From upper to lower: 1.0--1.6 keV, 1.6--2.2 keV,
2.2--2.8 keV, 2.8--3.4 keV, 3.4--5.6 keV, 5.6--7.8
keV. The HEG 1st-order events are used except below 1.6 keV, where the
MEG 1st-order data are utilized. The solid curve (red) represents the
best-fit continuum (modelled by an absorbed cutoff power law) modified
by absorption lines (negative gaussians) and a broad iron-K emission
(a positive gaussian). The upper axis of each panel marks 
wavelengths in units of \AA\ (blue).
\label{fig4}}
\end{figure*}

\tabcolsep=30pt
\begin{deluxetable}{cc}
\tablenum{1}
\tablecaption{Spectral Parameters of the Time-averaged HETGS Spectrum\label{tbl-1}}
\tablehead{\colhead{Parameter} &\colhead{Best-fit}}
\startdata
\cutinhead{Column Densities\tablenotemark{a} ($N_{\rm H}$)}
H ($10^{22}$ cm$^{-2}$)& 2.78$\pm$0.03 \nl
Mg ($10^{22}$ cm$^{-2}$)& 6.3$\pm$1.4 \nl
Si ($10^{22}$ cm$^{-2}$)& 5.8$\pm$0.4 \nl
S ($10^{22}$ cm$^{-2}$) & 7.6$\pm$1.2 \nl
Fe ($10^{22}$ cm$^{-2}$)\tablenotemark{b} & 10$\pm$1 \nl
\cutinhead{Continuum\tablenotemark{c}}
$A$ &7.15$\pm$0.25 \nl
$\Gamma$ &1.17$\pm$0.06 \nl
$E_{\rm fold}$ (keV) &3.80$\pm$0.17 \nl
\cutinhead{Iron-K Emission Line\tablenotemark{d}}
$E_{\rm cen}$ (keV)& 6.55$\pm$0.04\nl
1$\sigma$ width (keV) & 0.17$\pm$0.04\nl
E.W.            (eV) & 45$\pm$8 \nl
\hline
$E_{\rm cen}$   (keV) & 6.54$\pm$0.03 \nl
1$\sigma$ width (keV)& 0.05 (fixed)\nl
$r_{\rm in}$    ($r_{\rm g}$)&600$^{+300}_{-200}$ \nl
$r_{\rm out}$    ($r_{\rm g}$)& $100000$ (fixed)\nl
$\beta$ & $-3$ (fixed)\nl
\enddata
\tablenotetext{a}{Equivalent hydrogen column densities by assuming the solar
abundances \citep{and89} between the element and hydrogen (the abundance
ratios within each group of H-He-C-N-O, Ne-Na-Mg-Al, S-Cl-Ar-Ca, and
Cr-Fe-Co-Ni are fixed at the solar values). The columns of Mg, Si and
S are determined by the spectral features around each K edge, while that
of H is estimated from the fit to the whole HEG first order spectrum
by taking account of dust scattering based on the \citet{dra03} cross
section (see text).} 
\tablenotetext{b}{Determined from the \suzaku\ data.}
\tablenotetext{c}{A cutoff power law model with the form of
$A\;E^{-\Gamma} {\rm exp}(-E/E_{\rm fold})$ is adopted, where A is the
normalization at 1 keV in units of photons cm$^{-2}$ s$^{-1}$ keV$^{-1}$.}
\tablenotetext{d}{Modeled either by a single gaussian (upper section) or by a gaussian blurred with the {\it diskline} profile (lower).}
\tablecomments{The errors are 90\% confidence level for a single parameter.}
\end{deluxetable}

It is a critical issue to accurately determine the column densities of
major elements in the interstellar (plus circumstellar) matter toward
\grs\ for detailed study of its spectral model. In spectra with
insufficient energy resolution, a deep absorption edge produced by
interstellar matter can be coupled with other spectral features such
as emission lines at similar energies. It could also affect the
continuum shape at low energies where the photo-electronic absorption
cross section is large. The \chandra\ high resolution data provide us
with the best opportunity to solve this issue for this source.  The
result also has an important implication for study of elemental
abundances in the Galaxy, although we do not discuss it
here. \citet{lee02} determined column densities of Mg, Si, S, and Fe
by measuring the K-edge depth of each element, which are quite useful
and are adopted by some authors for spectral modelling
\citep[e.g.,][]{don04}.

In this paper, we re-determine those values from our data that have a
better statistics than that of \citet{lee02}, following the same
method described in \citet{ued05}. We adopt the photo-electric
absorption cross section by \citet{wil00}, which is available as {\it
TBvarabs} model in XSPEC.  The column density of lighter elements than
Mg is determined by a spectral fit to the whole continuum. Here we
consider the effect of dust scattering based on the cross section by
\citet{dra03}. Those of other elements are locally fit around the
K-edge by taking into account X-ray absorption fine structure (XAFS);
in particular, it is difficult to accurately estimate the column
density of Si just by measuring a depth of Si-K edge, due to the
complicated XAFS that is most likely attributed to silicates
grains. We thus model it using an experimental cross section of
SiO$_2$ as done in \citet{ued05}\footnote{empirically adding two
negative gaussians at 1.846 keV and 1.865 keV to a single absorption
edge at 1.839 keV works as a good approximation.}. Table~1 lists the
Hydrogen-equivalent column densities of each element assuming the
solar abundances by \citet{and89} for each element, which are used in
our subsequent analysis at their best-fit values. The relative
abundances within the groups of H-He-C-N-O, Ne-Na-Mg-Al, S-Cl-Ar-Ca,
and Cr-Fe-Co-Ni are fixed at the solar values. We find the column
density of Si is not as large as that estimated by \citet{lee02}. For
the column density of Fe, we adopt a value determined with \suzaku\
(Ueda, Y.\ et al.\ in preparation) because the \chandra\ result has a
large statistical error, yet consistent with the \suzaku\ result, due
to the limited energy coverage and area above 7.1 keV.

As seen from Figure~4, we detect 33 narrow absorption lines over the
whole energy range of HETGS, for which we simply model with negative
gaussians in the plot. We find that 32 of them correspond to
transition in highly ionized (H-like or He-like) ions. The absorption
line at 2.470 keV, identified as S~II, is most probably produced by
interstellar matter in the Galaxy, which is commonly seen in the high
resolution spectra of Galactic sources \citep{ued05}. The energy,
1$\sigma$ width, and equivalent widths are listed in Table~2 with its
identification. For weak lines whose width cannot be well determined,
we assume a velocity dispersion of 100 km s$^{-1}$ or 200 km s$^{-1}$,
depending on element. In addition to those narrow absorption lines, a
broad emission line feature centered at 6.55$\pm$0.04 keV is found,
attributable to K$\alpha$ lines of mildly ionized Fe ions. This can be
modeled by a single broad ($1\sigma$ width of 0.17$\pm$0.04 keV)
gaussian or by a narrower (0.05 keV) gaussian blurred with the ``disk
line'' profile \citep{fab89}, which is a more physical interpretation
for an iron-K line from the accretion disk (see \S~4.2 for
discussion). In the disk-line fit, we assume an outer radius of $10^5
r_{\rm g}$ with an emissivity law of $r^{\beta}$ where $r$ is the
radius and $\beta=-3$, and make the center energy and inner radius as
free parameters. The parameters of the emission line are summarized in
Table~1.

\tabcolsep=1pt
\begin{deluxetable}{ccccc}
\tabletypesize{\footnotesize}
\tablenum{2}
\tablecaption{List of Absorption Lines Observed in the HETGS spectra\label{tbl-2}}
 \tablehead{\colhead{$E_{\rm obs}$} &\colhead{$1\sigma$ width\tablenotemark{a}} &\colhead{E.W.} &\colhead{line ID\tablenotemark{b}} &\colhead{$E_{\rm theory}$ (term of final state)\tablenotemark{c}} \nl
\colhead{(keV)} &\colhead{(eV)} &\colhead{(eV)} &\colhead{} &\colhead{(keV)}}
\startdata

1.277(4) &0.43*        & 1.8&Ne~X 4p   &1.27707 ($^2P_{1/2}$), 1.27713 ($^2P_{3/2}$)\nl
1.4734(4)&0.001($<1.0$)& 1.7&Mg~XII 2p &1.47169 ($^2P_{1/2}$), 1.47264 ($^2P_{3/2}$)\nl
1.5522(6)&1.0(8)       & 1.4&Fe~XXIV 2s$\rightarrow$4p & 1.55091 ($^2P_{1/2}$), 1.55289 ($^2P_{3/2}$)\nl
\hline
1.7301(7)  &2(1) & 1.9&Al~XIII 2p                & 1.72769 ($^2P_{1/2}$), 1.72899 ($^2P_{3/2}$)\nl
     & & &Fe~XXIV 2s$\rightarrow$5p & 1.72876 ($^2P_{1/2}$), 1.72978 ($^2P_{3/2}$)\nl
1.7456(9)& 0.04($<1.1$)& 0.67&Mg~XII 3p                 & 1.74456 ($^2P_{1/2}$), 1.74484 ($^2P_{3/2}$)\nl
1.818(2)& 0.61*& 0.31&Ni~XXVI 2s$\rightarrow$4p & 1.81758 ($^2P_{1/2}$), 1.82033 ($^2P_{3/2}$)\nl
1.8276(5)& 0.61*& 0.75&Fe~XXIV 2s$\rightarrow$6p & 1.82640 ($^2P_{1/2}$), 1.82698 ($^2P_{3/2}$)\nl
1.842(2)& 0.61*& 0.54&Mg~XII 4p                 & 1.84002 ($^2P_{1/2}$), 1.84014 ($^2P_{3/2}$)\nl
1.8861(8)&0.63* & 0.49&Mg~XII 5p                 & 1.88419 ($^2P_{1/2}$), 1.88426 ($^2P_{3/2}$)\nl
2.0063(3)& 1.4(2)& 3.4&Si~XIV 2p                 & 2.00433 ($^2P_{1/2}$), 2.00608 ($^2P_{3/2}$)\nl
2.029(2)& 0.68*& 0.39&Ni~XXVI 2s$\rightarrow$5p & 2.02775 ($^2P_{1/2}$), 2.02916 ($^2P_{3/2}$)\nl
2.3777(5)&0.7($<1.8$)& 1.7&Si~XIV 3p                 & 2.37611 ($^2P_{1/2}$), 2.37663 ($^2P_{3/2}$)\nl
2.460(4)& 0.82*& 0.54&S~XV 2p                   & 2.46063 ($^1P_{1}$)\nl
2.470(1)& 2(1)& 1.3&S~II 3p\tablenotemark{d}     & 2.4694\tablenotemark{d}\nl
2.507(1)& 0.83*& 1.2&Si~XIV 4p                 & 2.50616 ($^2P_{1/2}$), 2.50638 ($^2P_{3/2}$)\nl
2.568(3)& 0.85*& 0.46&Si~XIV 5p                 & 2.56632 ($^2P_{1/2}$), 2.56644 ($^2P_{3/2}$)\nl
2.599(4)& 0.87*& 0.44&Si~XIV 6p                 & 2.59899 ($^2P_{1/2}$), 2.59906 ($^2P_{3/2}$)\nl
2.6229(3)& 2.5(3)& 5.2&S~XVI 2p                  & 2.61970 ($^2P_{1/2}$), 2.62270 ($^2P_{3/2}$)\nl
3.1077(7)& 0.9($^{+1.4}_{-0.8}$)& 2.0&S~XVI 3p                  & 3.10586 ($^2P_{1/2}$), 3.10675 ($^2P_{3/2}$)\nl
3.277(2)& 1.1*& 0.92&S~XVI 4p                  & 3.27589 ($^2P_{1/2}$), 3.27627 ($^2P_{3/2}$)\nl
3.3227(6)& 2.9(7)& 4.5&Ar~XVIII 2p               & 3.31818 ($^2P_{1/2}$), 3.32299 ($^2P_{3/2}$)\nl
3.355(4)& 1.1*& 0.44&S~XVI 5p                  & 3.35454 ($^2P_{1/2}$), 3.35473 ($^2P_{3/2}$)\nl
3.905(4)& 1.3*& 1.1&Ca~XIX 2p                 & 3.90226 ($^1P_{1}$)\nl
3.937(4)& 1.3*& 1.2&Ar~XVIII 3p               & 3.93429 ($^2P_{1/2}$), 3.93572 ($^2P_{3/2}$)\nl
4.1070(8)& 5(2)& 6.1&Ca~XX 2p                  & 4.10015 ($^2P_{1/2}$), 4.10750 ($^2P_{3/2}$)\nl
4.151(6)& 1.4*& 0.80&Ar~XVIII 4p               & 4.14974 ($^2P_{1/2}$), 4.15034 ($^2P_{3/2}$)\nl
4.865(5)& 1.6*& 1.7&Ca~XX 3p                  & 4.86192 ($^2P_{1/2}$), 4.86410 ($^2P_{3/2}$)\nl
5.684(6)& 3.8**& 3.3&Cr~XXIII 2p               & 5.68205 ($^1P_{1}$)\nl
5.931(4)& 7($<16$)& 5.8&Cr~XXIV 2p                & 5.91650 ($^2P_{1/2}$), 5.93185 ($^2P_{3/2}$)\nl
6.17(2) & 4.1**& 3.6&Mn~XXIV 2p                & 6.18044 ($^1P_{1}$)\nl
6.45(1) & 17($^{+14}_{-12}$)& 7.9&Mn~XXV 2p                 & 6.42356 ($^2P_{1/2}$), 6.44166 ($^2P_{3/2}$)\nl
6.692(3) & 28(3)& 40&Fe~XXV 2p                 & 6.70041 ($^1P_{1}$)\nl
     & & &Fe~XXIV 2p                 & 6.67644 ($^2P_{1/2}$), 6.67915 ($^2P_{3/2}$)\nl
6.975(2) & 16(2)& 40&Fe~XXVI 2p                 & 6.95196 ($^2P_{1/2}$), 6.97317 ($^2P_{3/2}$)\nl
     & & &Cr~XXIV 3p                 & 7.01726 ($^2P_{1/2}$), 7.02181 ($^2P_{3/2}$)\nl
\enddata
\tablenotetext{a}{*, **: fixed at a value corresponding to a velocity dispersion of 100 km/s (*) and 200 km/s (**).}
\tablenotetext{b}{The initial state is ``1s'' unless otherwise indicated.}
\tablenotetext{c}{Taken from the NIST database (version 3.1.5) except for the S~II 3p line,
 for which an observed value by \citet{ued05} is shown.}
\tablenotetext{d}{Most probably due to the interstellar gas (see text).}
\tablecomments{The MEG data are utilized for the absorption lines below 1.6 keV.}
\tablecomments{The single number in the parenthesis indicates the statistical error in the last digit (90\% confidence level for a single parameter).}
\end{deluxetable}

To determine the physical parameters of the ionized plasma, we next
fit the absorption line profile of major elements (Si, S, Ar, Ca, Cr,
Mn, Fe) by a Voigt function in the local energy band. The Voigt
profile, whose opacity is expressed by a convolution of a gaussian and
a lorenzian (natural broadening), is a physically more correct model
rather than a negative gaussian. To do this, we utilize a local model
(called ``Kabs'') implemented on XSPEC as described in \citet{ued04},
where its full formula is given. The free parameters are three per
each ion, a column density (cm$^{-2}$), a line-of-sight velocity
dispersion $b$ (km s$^{-1}$), and a Doppler shift $z = v/c$, where $v$
is the mean bulk velocity of plasma (negative for outflow) and $c$ is
the light velocity.  The atomic data are taken from the database
provided by National Institute of Standards and Technology
(NIST)\footnote{Version 3.1.5, http://physics.nist.gov/asd3 [2008
September 6]}. In this paper, we treat an ideal case where
contribution of re-emission line from the plasma out of the
line-of-sight is neglected, corresponding to Case~I in \citet{ued04}
or ``non-radiative de-excitation limit'' in \citet{mas04}. 
It is because the contribution of the emission cannot be well
constrained by our data, being consistent with zero within the
statistical errors; when we include such emission lines emitted at $r=
(1-3)\times 10^{5} r_{\rm g}$ with an emissivity law of $r^{-3}$
(corresponding to the case of an optically thin, self-similar 
flow with a constant velocity) and an inclination of
70$^\circ$ in the spectral model, we obtain an upper limit for the
solid angle of the plasma as $\Omega / 2\pi < 0.11$ (90\% confidence
level) from a simultaneous fit to the Si XIV and S XVI absorption
lines. We find that inclusion of these emission lines at the maximum
level could increase the column density obtained for the same velocity
dispersion by $\approx$30\% for Si XIV and $\approx$50\% for Fe XXVI,
which are within the statistical errors.

We hereafter analyze the HETGS spectra of Epochs I and II separately,
since the absorption line profile could change according to the
evolution of the continuum spectrum. Figure~5 shows the HEG 1st-order
spectra corrected for effective area (i.e., incident spectra except
for blurring with the energy resolution). It is evident that the
spectrum is harder in Epoch II than in Epoch I. In this subsection, we
focus on the absorption lines. Detailed analysis of the continuum is
given in next subsection using the \rxte\ PCA data, which have a wider
energy coverage than the HETGS.

\begin{figure}
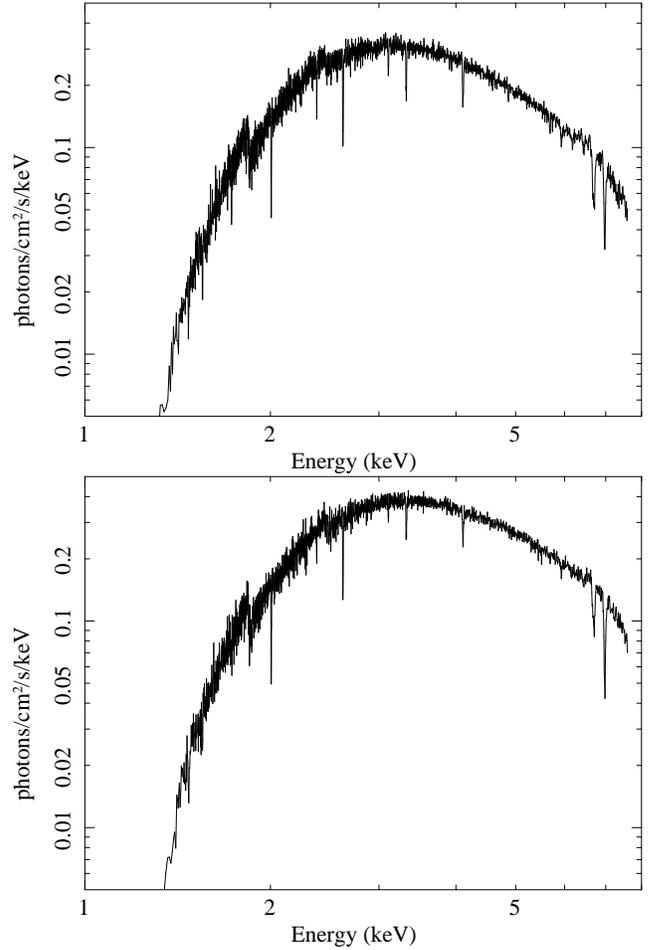

\epsscale{1.0}
\includegraphics[angle=270,scale=0.37]{f5a.ps}
\includegraphics[angle=270,scale=0.37]{f5b.ps}
\caption{
Upper: The HEG 1st-order spectrum (corrected for effective area) in Epoch~I.
Lower: The HEG 1st-order spectrum (corrected for effective area) in Epoch~II.
\label{fig5}}
\end{figure}

To best constrain the physical parameters of each ion, we fit two
absorption lines of K$\alpha$ and K$\beta$, whenever available (i.e.,
except for Mn and Fe), simultaneously in the local bands. This works
quite well to de-couple the inevitable anti-correlation between a
column density and a velocity dispersion in the curve-of-growth
analysis. We simply adopt an absorbed cut-off power model for the
continuum by fixing $E_{\rm fold}$ and the interstellar
absorption. The velocity dispersion in two epochs is found to be
consistent with a constant value within the errors. We thus assume the
best-fit velocity dispersion obtained from the summed spectrum of
Epochs I and II when deriving the column density and Doppler shift of
the Si, S, Ar, and Ca ions in each epoch. In the iron-K band, we
simultaneously fit K$\alpha$ and K$\beta$ lines of Cr XXIV, K$\alpha$
of Mn XXV, and K$\alpha$ of Fe XXVI by assuming a same velocity
dispersion and Doppler shift. The Cr XXIV K$\beta$ line (7.02 keV)
contaminates a high energy part of the Fe XXVI K$\alpha$ feature in
the HETGS spectra. We treat the Fe XXV absorption line independently,
considering possible uncertainties in the incident line
energies\footnote{Unlike a single electron system in H-like ions,
there could be more uncertainties in the atomic database for He-like
(and more electrons) ions.} and a contamination of less ionized ions
such as Fe XXIV, although we assume the same velocity dispersion as
that of Cr XXIV, Mn XXV, and Fe XXVI. 
We note that the results of Fe XXV may not be reliable to be taken at
their face values but just provide a rough estimate. For Si XIV and Fe
XXVI, we introduce an upper limit for the column density constrained
by the corresponding K-edge depth measured by HETGS and by PCA,
respectively; otherwise, the fit to the absorption line profile leads
to unphysically large column densities.

Figure~6 shows a blow-up of absorption line profile in Epochs I and II
for each set of elements around their K$\alpha$ lines. The solid
curves represent the best-fit model. Table~3 gives the summary of the
parameters. We find that all the ions have significant blue shifts,
$\approx$150 km s$^{-1}$ for Si XIV and $\approx$500 km s$^{-1}$ for
Fe XXVI. The difference of the velocity between H-like Si (and S, Ar,
Ca) and Fe ions is robust, larger than the systematic error in the
relative wavelength calibration as mentioned in \S~2.1.  The velocity
dispersion increases with the atomic number, from 70 km s$^{-1}$ for
Si to $\approx$ 200 km s$^{-1}$ for Fe. From Epoch I to II, the column
density of all the ions decreases, indicating that the plasma was in a
higher ionization stage in the latter epoch than the former; the
number ratio between Fe XXV and Fe XXVI is also consistent with this
picture.

\begin{figure*}
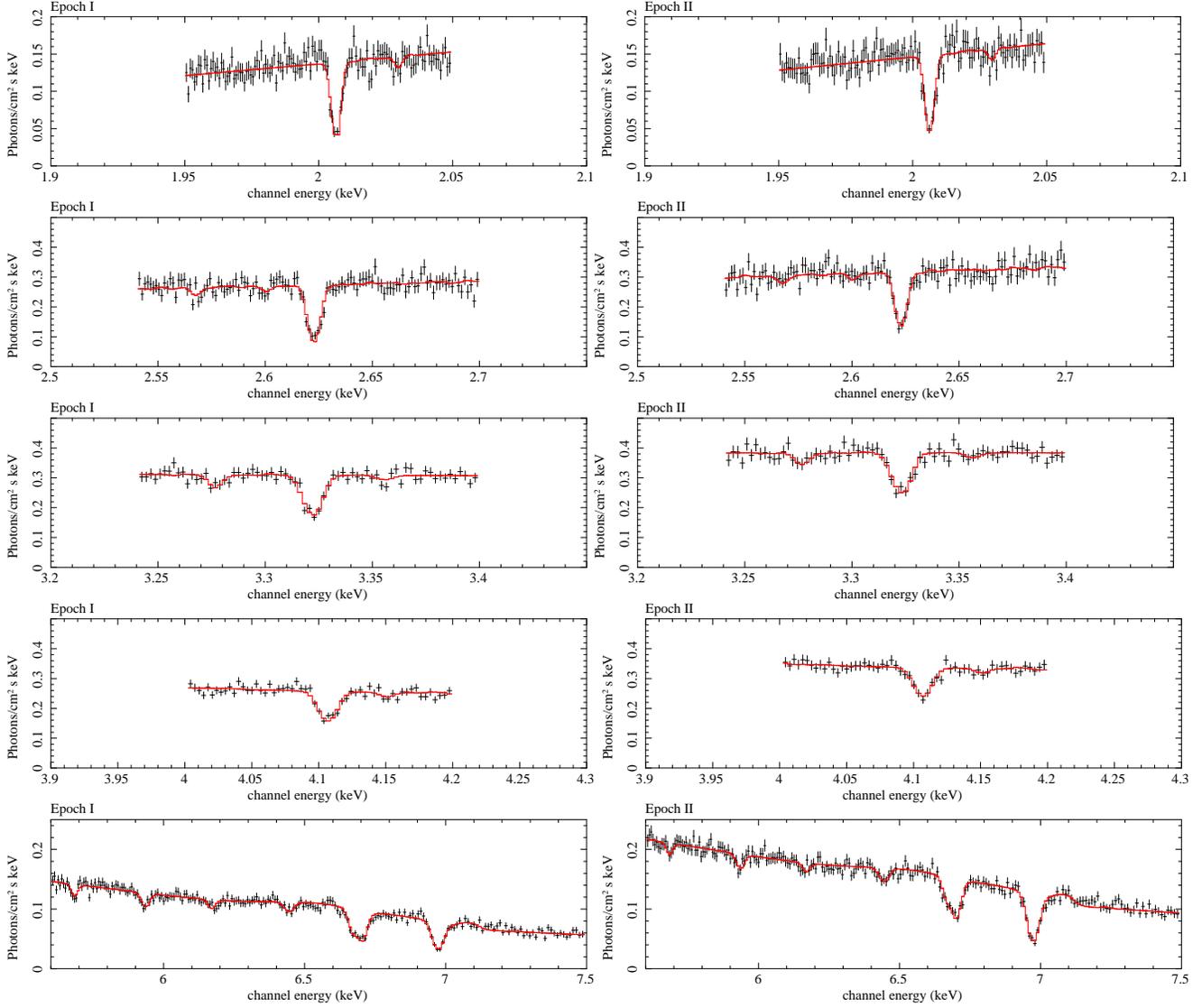

\epsscale{1.0}
\includegraphics[angle=270,scale=0.37]{f6a.ps}
\includegraphics[angle=270,scale=0.37]{f6b.ps}

\includegraphics[angle=270,scale=0.37]{f6c.ps}
\includegraphics[angle=270,scale=0.37]{f6d.ps}

\includegraphics[angle=270,scale=0.37]{f6e.ps}
\includegraphics[angle=270,scale=0.37]{f6f.ps}

\includegraphics[angle=270,scale=0.37]{f6g.ps}
\includegraphics[angle=270,scale=0.37]{f6h.ps}

\includegraphics[angle=270,scale=0.37]{f6i.ps}
\includegraphics[angle=270,scale=0.37]{f6j.ps}
\caption{
A blowup of the time-averaged HEG 1st-order spectra
around each K-absorption line in Epochs I (left) and II (right). 
The best-fit model employing Voigt functions is plotted with a line (red).
From upper to lower: Si XIV, S XVI, Ar XVIII, Ca XX, 
Cr XXIV + Mn XXV + Fe XXV + Fe XXVI.
\label{fig6}}
\end{figure*}

\tabcolsep=5pt
\begin{deluxetable*}{cccccccc}
\tabletypesize{\footnotesize}
\tablenum{3}
\tablecaption{Parameters of the K$\alpha$ Absorption Lines Fit by Voigt Profile\label{tbl-3}}
\tablehead{
\colhead{}
&\multicolumn{3}{c}{Epoch I}&\colhead{}
&\multicolumn{3}{c}{Epoch II}\\
\cline{2-4} \cline{6-8}\\
\colhead{Ions}
&\colhead{$-z$\tablenotemark{a}} &\colhead{$b$\tablenotemark{b}} &\colhead{$N_{\rm ion}$} 
&\colhead{}
&\colhead{$-z$\tablenotemark{a}} &\colhead{$b$\tablenotemark{b}} &\colhead{$N_{\rm ion}$ }\\
\colhead{}
&\colhead{($\times10^{-3}$)} &\colhead{(km s$^{-1}$)} &\colhead{($10^{18}$ cm$^{-2}$)}
&\colhead{}
&\colhead{($\times10^{-3}$)} &\colhead{(km s$^{-1}$)} &\colhead{($10^{18}$ cm$^{-2}$)}
}
\startdata
Fe XXVI
& $1.68\pm0.18$ & 190$^{+40}_{-30}$ & 13$^{+7}_{-8}$  
&& $1.72^{+0.09}_{-0.05}$ & 210$\pm 10$ & 10$^{+0}_{-3}$  \nl
Fe XXV\tablenotemark{c} 
& $0.66\pm0.07$ & (= Fe XXVI) & 7.2$\pm1.4$  
&& $0.31\pm0.08$ & (= Fe XXVI) & 2.2$\pm0.3$  \nl
Mn XXV 
& (= Fe XXVI) & (= Fe XXVI) & 0.19$\pm0.06$  
&& (= Fe XXVI) & (= Fe XXVI) & 0.16$\pm0.06$  \nl
Cr XXIV 
& (= Fe XXVI) & (= Fe XXVI) & 0.19$\pm0.04$  
&& (= Fe XXVI) & (= Fe XXVI) & 0.12$\pm0.05$   \nl
Ca XX 
& $0.61^{+0.13}_{-0.19}$ & 110$^{+300}_{-20}$* & 0.31$^{+0.12}_{-0.06}$  
&& $0.63^{+0.17}_{-0.54}$ & (= Epoch I)* & 0.18$^{+0.04}_{-0.05}$   \nl
Ar XVIII
& $0.41^{+0.10}_{-0.13}$ & 110$^{+170}_{-40}$* & 0.26$\pm 0.05$  
&& $0.62^{+0.20}_{-0.15}$ & (= Epoch I)* & 0.14$\pm 0.02$   \nl
S XVI
& $0.51^{+0.09}_{-0.03}$ & 105$^{+4}_{-9}$* & 0.47$^{+0.08}_{-0.09}$  
&& $0.55^{+0.08}_{-0.13}$ & (= Epoch I)* & 0.23$^{+0.07}_{-0.03}$   \nl
Si XIV
& $0.55^{+0.07}_{-0.04}$ & 74$^{+10}_{-3}$* & 0.47$^{+0.16}_{-0.11}$  
&& $0.50\pm0.06$ & (= Epoch I)* & 0.35$^{+0.10}_{-0.08}$   \nl
\enddata
\tablenotetext{a}{Blue shifts (i.e., outflow velocity relative to the light velocity).}
\tablenotetext{b}{Line-of-sight velocity dispersion.}
\tablenotetext{c}{There may be systematic uncertainties in the 
parameters of Fe XXV (see text).}
\tablecomments{* best-fit parameter determined by the time-averaged spectrum.}
\tablecomments{The errors are 90\% confidence level for a single parameter.}
\end{deluxetable*}

\subsection{Continuum Model}

Once the local spectral features and interstellar absorption are
accurately determined with the \chandra\ data, we can now study the
continuum emission with the \rxte\ PCA spectra in the 3--25 keV range
with least uncertainties even if they cannot be resolved with the PCA
data alone. We assume that the PCA spectra integrated in Epochs I' and
II' well represent the HETGS spectra in Epochs I and II, respectively,
although the time coverage by \rxte\ is shorter than that by
\chandra. In the analysis of PCA spectra, we fix the parameters of
absorption/emission lines and absorption column densities at the the
best-fit values obtained above. We introduce four additional spectral
features that reside outside the HETGS band: absorption lines at 7.84
keV and 8.20 keV and edge structures at 8.6 keV and 9.0 keV. The
absorption lines correspond to Fe XXV K$\beta$ + Ni XXVII K$\alpha$
and Fe XXVI K$\beta$ + Ni XXVIII K$\alpha$, for which we assume an
equivalent width of 30 eV and 20 eV, respectively \citep{kot00}. The
edge models approximately represent spectral modification by the
bound-free (photo-absorption) and bound-bound transition close to the
K-edge energy of Fe XXV and Fe XXVI ions, respectively. The ratio of
the edge depth between Fe XXV and Fe XXVI is fixed at 1.1 (Epoch~I')
and 0.44 (Epoch~II'), based on the results from the HETGS. In the
continuum modeling, we always include a reflection component that is
expected from the broad iron-K emission line detected with the
HETGS. We model it with the {\it pexriv} code \citep{mag95}, or its
convolution model. We fix the inclination, temperature, abundances of
Fe, and that of other metals to be 70$^\circ$, $10^6$ K, 3.6 solar,
and 2.1 solar, respectively. To make it consistent with the central
energy and equivalent width of the iron-K emission line observed with
the HETGS, the ionization parameter $\xi_{\rm refl}$ is fixed at 20
(Epoch I') and 40 (Epoch II'), and the solid angle of the reflector
$\Omega_{\rm refl}/2\pi$ is restricted be less than 0.5. Since we
consider the accretion disk as the reflector, it is further blurred
with the {\it diskline} kernel assuming an emissivity profile of
$r^{-3}$ from radii of $(600-10^5) r_{\rm g}$, again based on the
results of the HETGS for the iron-K line.

We find that a cutoff power law model give a good (mathematical)
description for the overall PCA spectra in the 3--25 keV range as is
for the HETGS spectra. As a physical model, we first try the multi
color disk (MCD) model \citep{mit84} plus a power law, which has been
widely used to fit the spectra of canonical black holes in the
high/soft state \citep[see e.g.,][]{tan95}. This model is not
acceptable for our PCA spectra of \grs, with reduced $\chi^2$ (d.o.f)
of 2.19 (44) and 3.98 (44) for Epoch I' and II', respectively. Note
that this model can however fit the HETGS spectra whose coverage is
limited below 7.5 keV, demonstrating that continuum modeling only by
using data below $<$10 keV are sometimes unreliable. Next, we try a
power law plus a ``$p$-free disk'' model, where the temperature
profile is express as a $T = T_{\rm in} (r/r_{\rm in})^{-p}$ with $p$
being a free parameter instead of a fixed value of 0.75 in the MCD
model. This model has been successfully applied for black holes
accreting at high fractional Eddington ratio
\citep[e.g.,][]{min94,kub04a}. We find that the fit is significantly
improved (and is even successful for the Epoch~II' data); we obtain
$p=0.43^{+0.3}_{-0.2}$, $T_{\rm in}=2.31\pm 0.12$ keV, $r_{\rm
in}=5.7^{+2.1}_{-0.7}$ km with reduced $\chi^2$ (d.o.f) of 1.46 (43)
for Epoch I', and $p=0.49^{+0.3}_{-0.2}$, $T_{\rm in}= 2.48\pm0.09$
keV, and $r_{\rm in}= 6.1^{+1.7}_{-0.6}$ km with reduced $\chi^2$
(d.o.f) of 1.01 (43) for Epoch II'. Here $r_{\rm in}$ is corrected for
cosine of the inclination by assuming the emission region is more like
a disk rather than spherical geometry.

We finally try a Comptonization model, which has been used to model
the spectra of black holes in the very high state
\citep[e.g.,][]{kub04a} and the low/hard state
\citep[e.g.,][]{gir97}. This model is found to be a good description
of the spectra of \grs\ in several cases \citep[][]{don04}. Here we
utilize {\it thComp} code \citep{zyc99}, which is appropriate even
when the optical depth of Comptonizing corona is larger than 3, as in
our case. The spectrum of seed photons is assumed to be the MCD
model. The free parameters are the temperature of Comptonizing
electrons $T_{e}$, the photon index $\Gamma$, with the innermost
temperature $T_{\rm in}$ and radius $r_{\rm in}$ for the incident MCD
component. We can infer $r_{\rm in}$ from the conservation of photon
numbers in the Comptonization process (see \S~4.2). The scattering
optical depth can be calculated from $T_{\rm e}$ and $\Gamma$ according to
the following formula \citep{sun80}:
$$\tau_{\rm e} = \sqrt{2.25+\frac{3}{(T_{\rm e}/511\;{\rm keV})
    \cdot [(\Gamma+0.5)^2-2.25]}} - 1.5.
$$ 
Since we find no direct (i.e., without Comptonization) MCD component,
we do not included it in our model, unlike the case of \gro\ \citep{tak08} and
Cygnus~X-1 \citep{mak08}. The upper limit on the fraction of the
non-Comptonized flux to the Comptonized one (the flux before being
Comptonized) is $20\%$ (Epoch I') and $3\%$ (Epoch II).

The fit with the Comptonization model is found to be much better than
the $p$-free disk model, yielding reduced $\chi^2$ (d.o.f) of 0.94
(43) and 0.56 (43) for Epoch I' and II', respectively. Figure~7
(upper) show the PCA spectra folded with the energy response, with the
best-fit model (solid line) and residuals in units of
$\chi$. Figure~7 (lower) show the incident photon spectra in the form
of $E^2 F_E$, where $E$ and $F_E$ is the energy and photon spectrum,
respectively. The contribution of the reflection component including
an iron-K emission line calculated self-consistently is separately
plotted. The best-fit parameters (including those of the reflection
component) are summarized in Table~4. We find the Comptonizing corona
has a low electron temperature of $\approx$3.6 keV and a large
scattering optical-depth of $\approx 5$, which are constant in the two
epochs within the error. The hardening in Epoch II can be attributed
to the intrinsic change of the seed-photon spectrum. The 0.01--100 keV
intrinsic luminosity corrected for absorption and reflection is found
to be $6.6\times10^{38}$ erg s$^{-1}$ (Epoch I') and
$8.3\times10^{38}$ erg s$^{-1}$ (Epoch II'), by assuming isotropic
emission. These correspond to 0.3--0.4 times the Eddington limit of 14
\solarmass\ black hole, and could become 0.9--1.2 times the Eddington
if the emission region is optically thick and has a disk-like
geometry. We discuss the structure of the accretion disk in \S~4.2.

\begin{figure*}
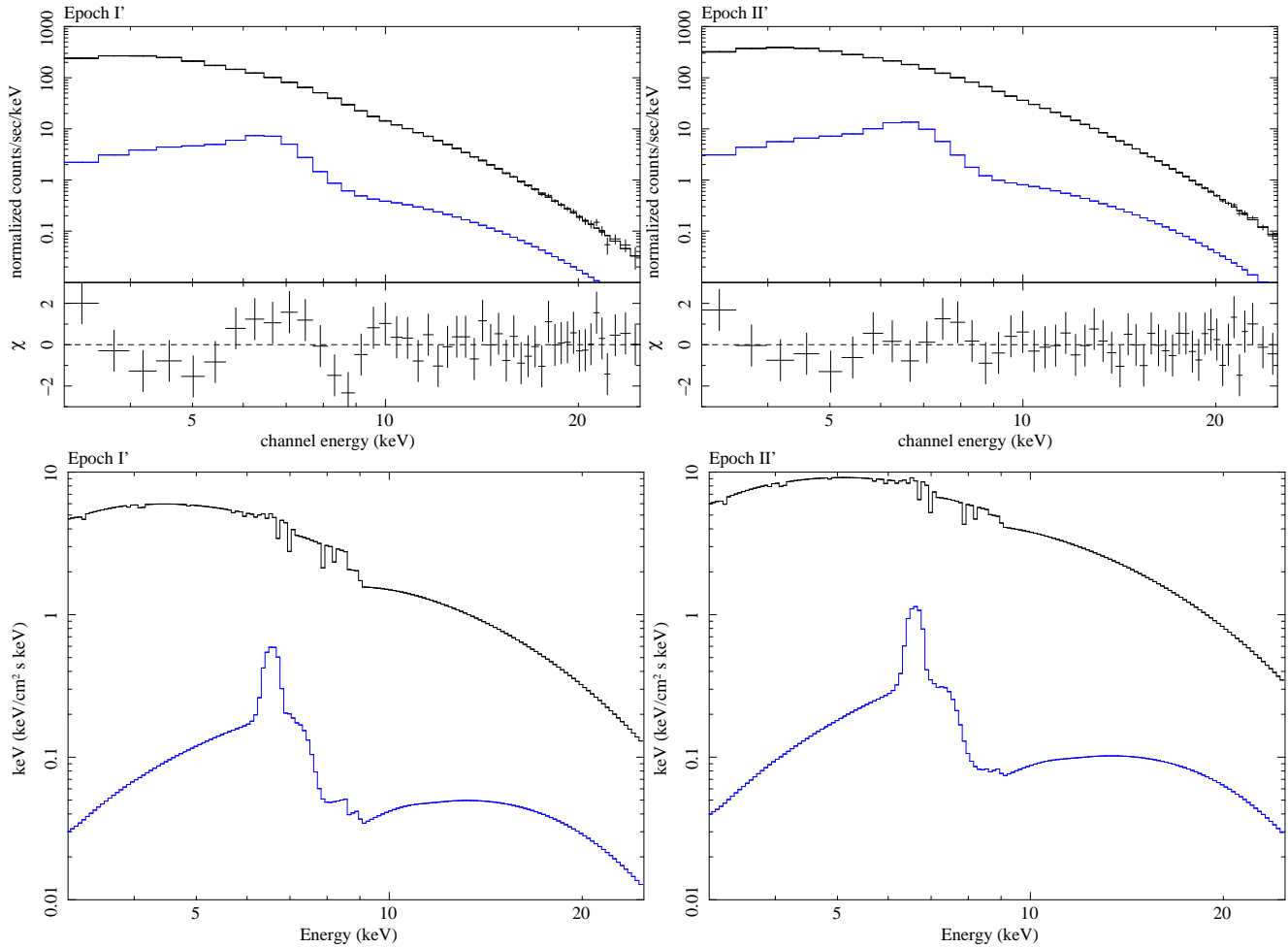

\epsscale{1.0}
\includegraphics[angle=270,scale=0.37]{f7a.ps}
\includegraphics[angle=270,scale=0.37]{f7b.ps}

\includegraphics[angle=270,scale=0.37]{f7c.ps}
\includegraphics[angle=270,scale=0.37]{f7d.ps}
\caption{
Upper: The \rxte\ PCA spectra (with error bars) folded with the energy
response in Epoch I' (left) and II' (right). The upper curve (black)
represents the best-fit model including the contribution of the
reflection component (lower curve, blue). The lower panel shows the
residuals of the data from the best-fit model, normalized by the error
(i.e., in units of $\chi$). Lower: the best-fit model in the form of
$E^2 F_E$, where $E$ and $F_E$ is the energy and photon spectrum,
respectively, in Epoch I' (left) and Epoch II'' (right). The lower
curve (blue) represents the reflection component.
\label{fig7}}
\end{figure*}

\tabcolsep=5pt
\begin{deluxetable}{ccccc}
\tablenum{4}
\tablecaption{Best-Fit Parameters of the \rxte\ PCA Spectra\label{tbl-4}}
\tablehead{\colhead{Parameter} &\colhead{Unit} &\colhead{} &\colhead{Epoch I'} &\colhead{Epoch II'}}
\startdata
$r_{\rm in}$ &(km) && 113$\pm$7& 79$\pm$8 \nl
$T_{\rm in}$ &(keV) && 0.87$\pm$0.06 & 1.11$\pm$0.06 \nl
$T_{\rm e}$ &(keV) && 3.62$^{+0.52}_{-0.24}$ & 3.66$^{+0.34}_{-0.25}$ \nl
$\Gamma$ &&& 3.18$^{+0.17}_{-0.15}$ & 2.99$\pm$0.14\nl
( $\tau_{\rm e}$\tablenotemark{a} &&& 4.8$\pm$0.3& 5.2$\pm$0.4 ) \nl
$\Omega_{\rm refl}/2\pi$\tablenotemark{b} &&& 0.5$^{+0.0}_{-0.06}$ & 0.42$^{+0.08}_{-0.19}$ \nl
$\xi_{\rm refl}$\tablenotemark{b} &(erg cm s$^{-1}$)&& 20 (fixed)& 40 (fixed)\nl
\hline
$\tau_{\rm edge}$ (9.0 keV)\tablenotemark{c} &&& 0.25$\pm0.02$& 0.16$\pm$0.03\nl
$\tau_{\rm edge}$ (8.6 keV)\tablenotemark{c} &&& 0.28$\pm0.03$& 0.07$\pm$0.02\nl
\hline
$F_{2-10}$\tablenotemark{d} &(erg cm$^{-2}$ s$^{-1}$) &&$1.0\times10^{-8}$ & $1.6\times10^{-8}$ \nl
$L_{0.01-100}$\tablenotemark{e} &(erg s$^{-1}$)&& $6.6\times10^{38}$ & $8.3\times10^{38}$ \nl
\hline
$\chi^2$ (d.o.f)\tablenotemark{f} &&& 40.4 (43)& 24.1 (43)\nl
\enddata
\tablenotetext{a}{The optical depth of the Compton cloud calculated
from $T_{\rm e}$ and $\Gamma$ (see text).}
\tablenotetext{b}{A reflection component is included with a solid angle 
$\Omega_{\rm refl}$ and an ionization parameter $\xi_{\rm refl}$, blurred by the {\it diskline}
kernel assuming the emissivity profile of $r^{-3}$ between
$r_{\rm in} = 600\; r_{\rm g}$ and $r_{\rm out} = 10^5\; r_{\rm g}$.}
\tablenotetext{c}{Optical depth of the {\it edge} model at 9.0 keV and
8.6 keV.  The ratio between 9.0 and 8.6 keV is fixed at 1.1 (Epoch I')
and 0.44 (Epoch II').}
\tablenotetext{d}{Observed flux in the 2--10 keV band.}
\tablenotetext{e}{Intrinsic luminosity of the direct continuum
(without the reflection component) in the 0.1--100 keV band, corrected
for absorption. The Comptonized radiation is assumed to be isotropic.}  
\tablenotetext{f}{The $\chi^2$ value and degree of freedom.}
\tablecomments{The errors are 90\% confidence level for a single parameter.}
\end{deluxetable}

\section{Discussion}

\subsection{Nature of Accretion Disk Wind}

We have detected at least 32 narrow absorption lines from a highly
ionized plasma (H-like or He-like) in the HETGS spectra of \grs\ in
``soft state'' that are the deepest among those detected from the
source, achieving our primary aim with this ToO observation. This
enables us to investigate the physical properties of the plasma with
the best accuracy ever achieved. The velocity and line-of-sight
velocity dispersion are determined to be 150--500 km s$^{-1}$ and
70--200 km s$^{-1}$, respectively, which differ by element. Since the
proper motion of \grs\ is estimated at $3\pm10$ km s$^{-1}$ from
\citet{gre01} (considering the fact that the black hole mass is about
12 times larger than the companion in this system), it is revealed
that the plasma is definitely in outflow, and hence the origin is 
a wind (most likely an accretion disk wind) as suggested in previous
studies \citep{lee02}. The depth of these absorption lines changed
according to the X-ray flux, consistent with the picture that it is
photo-ionized by the radiation.

Estimating the location of the wind is crucial to discriminate its
origin. We estimate it by a simple argument using the ionization
parameter $\xi = L/nr^2$ \citep{tar69}, where $L$ is the luminosity,
$n$ is the density, and $r$ is the distance from the source. As
discussed below, $\xi$ may not always constant along the line of sight
but here we discuss its averaged value for first order estimate. 
For
this purpose, we perform simulation of a photoionized plasma by the
XSTAR code for various values of $\xi$, assuming our best-fit
continuum models in Table~4 and the metal abundances in Table~1. The
density $n$ is set to be $10^{12}$ cm$^{-3}$.  We find that the
observed number ratio between Fe XXVI and Fe XXV ions, 1.8$\pm$1.0 and
4.5$\pm$1.5, indicates log $\xi = 4.3^{+0.1}_{-0.6}$ and log $\xi =
4.2\pm0.1$ for Epoch I and II, respectively; here $L$ is defined in
the 0.0136--13.6 keV band in each spectrum (this is the reason why the
$\xi$ value is larger in Epoch~I in spite of the lower ionization
state). Assuming that the length of the wind, $\Delta r$, is a similar
order to the distance from the source, we find $n r \sim n \Delta r
\sim 10^{23}$ cm$^{-2}$ after correcting for the fraction of Fe XXVII
ions and Fe abundance of 3.6 times the solar. This yields $r \sim L /
\xi / nr \sim (2-6)\times10^{11}$ cm, or $\sim (1-3)\times 10^5\;
r_{\rm g}$ for a black hole mass of 14 \solarmass\footnote{
Note that here we have assumed that the gas is smoothly
distributed i.e., that it is not clumped. Clumping would reduce the
radius, and hence this estimate is an upper limit.}.
This well exceeds a
possible minimum launching point of thermally driven wind from an
X-ray irradiated disk theoretically predicted. The critical radius
where the thermal energy of the X-ray heated disk exceeds the
gravitational energy is given by
$$
r_{\rm c} \sim 10^{10} (\frac{10^8\; {\rm K}}{T_{\rm C}}) (\frac{M}{\solarmass}) \;\;{\rm cm}.
$$
Here $T_{\rm C}$ is the Compton temperature, at which the net energy
transfer between electrons and photons is balanced out.
In our case $T_{\rm C}$ is estimated to be $\sim$0.3 keV by taking
the average of photon energy of the continuum in the 0.01--100 keV band,
and hence $r_{\rm c} \sim 4\times10^{12}$ cm. It is predicted
that such wind can be launched from $\sim 0.1 r_{\rm c}$ or even much
smaller radii \citep{woo96}. As the luminosity ($>6.6\times10^{38}$
erg s$^{-1}$ in the 0.1--100 keV band) is about 0.3 times the
Eddington limit, the effect of radiation should be also important even
if the gas is highly ionized and the opacity by UV-lines is
little. Thus, the wind we observe is consistent with a thermally (plus
radiation) driven wind and there is no necessity to invoke a
magnetically driven wind. 

The mass outflow rate carried by the accretion disk wind is evaluated as 
\begin{eqnarray*}
\dot{M}_{\rm out} &=& \Omega_{\rm wind} r^2 n m_{\rm p} v_{\rm wind}\\
      &\sim& (0.6-1.3)\times10^{19} \; (\frac{\Omega_{\rm wind}/4\pi}{0.2}) (\frac{v_{\rm wind}}{500\; {\rm km \; s^{-1}}})\;\; {\rm g \; s^{-1}}.
\end{eqnarray*}
Here we have used $r^2 n = L/\xi \sim (3-6)\times10^{34}$, $m_{\rm
  p}$ is the proton mass, $\Omega_{\rm wind}$ is the solid angle of
the wind, and $v_{\rm wind}$ is the outflow velocity (which is larger
than the observed outflow velocity by a factor of cosine of the angle
between the direction of the wind and our line-of-sight). Thus, the
mass {\it outflow} rate carried by the wind could be comparable to 
the mass {\it accretion} rate in the inner part of the disk, which is
estimated to be $(0.7-0.9)\times 10^{19}$ g s$^{-1}$ (spherical case) or
$(2-3)\times 10^{19}$ g s$^{-1}$ (disk-like case) by assuming the
mass-to-energy conversion efficiency of 0.1. This demonstrates the
importance of the wind for understanding the dynamics of the whole
disk system. A similar situation is found in the neutron star low mass
X-ray binary GX 13+1 \citep{ued04} except for the difference of the mass
scale.

We find that H-like ions of heavier elements have higher outflow
velocity and larger velocity dispersion. In addition, there is an
implication that Fe XXV shows a smaller velocity than that of Fe XXVI, although
we have to keep in mind the possible systematic uncertainties in the
obtained parameters for Fe XXV. These results indicate that we trace
different parts of the wind by different ions; there is a component of
the flow that is more highly ionized (i.e., traced by Fe XXVI) and
that is moving approximately 3 times faster than the gas traced by
other lines. According to a thermally driven wind model
\citep[see][and references therein]{net06}, the velocity of outflow
can be represented in a power law form of distance,
$$
v \propto r^{\gamma},
$$
where $\gamma$ is in the range of 0--0.5. 
Then, from the conservation of the mass outflow rate, 
the density is given as 
$$
n \propto r^{-2-\gamma},
$$
and therefore
$$
\xi \propto r^{\gamma}.
$$
For $\gamma > 0$, ionization becomes stronger at outer parts of the
wind. In this interpretation, H-like Si ions we observe must be
located closer to the center where $\xi$ is smaller, and hence have a
smaller outflow velocity, compared with those of H-like Fe ions.
The larger velocity dispersion found in heavier element of H-like ions
can be attributed to a larger line-of-sight velocity gradient when
integrated in a more outer region of the wind, probably accompanied by
larger turbulent motion as well. Constructing a fully quantitative
model for the wind of \grs\ is beyond the scope of this paper and is
left for future studies.

\subsection{Origin of X-ray Continuum Emission in ``Soft State''}

The ideal combination of the \chandra\ HETGS and \rxte\ PCA has
enabled us to constrain the continuum model with least uncertainties,
by correctly taking into account the interstellar absorption with the
(updated) anomalous abundances and local spectral features including
absorption- and emission lines. The spectra of \grs\ in ``soft state''
(steady State~A in the definition of \citealt{bel00}) in the 1--25 keV
band can be well explained with a strongly Comptonized continuum and
its reflection component by an outer part of the disk, which is
absorbed by a highly ionized disk wind discussed above.

A relativistically broadened iron-K emission line as previously
reported by \citet{mar02} is not required in our data, although it is
difficult to perfectly exclude its presence from the spectral fit
alone, which strongly depends on the continuum model (and also on the
interstellar absorption). The large inner radius we obtained from the
line profile with the \chandra\ HETGS, $400 r_{\rm g}$, does not
necessarily mean that the accretion disk is truncated at that
radii. Rather, as discussed below, due to the strong Comptonization by
corona that completely covers the inner disk, any iron-K features
arising there should be smeared out. This makes extremely difficult to
constrain the innermost radius of the accretion disk from the iron-K
emission profile.

We find that a thermal Comptonization is the most likely origin of the
continuum emission of \grs\ in ``soft state''. The standard, MCD plus
a power law description for the continuum does not give acceptable
fit. Instead of the MCD model, the $p$-free disk model, expected from
a one-dimensional slim disk \citep{abr88}, gives a statistically
acceptable fit to the spectrum in Epoch~II'. We consider it physically
unlikely, however, because the inferred innermost radius is too small
($ \simlt 0.3 r_{\rm g}$) than a theoretical prediction
\citep{wat00}. Also, we cannot consistently explain the spectra of two
epochs with the same model.

The spectral model of Comptonization yields an optical depth $\approx
5$ and a low electron temperature ($\approx $4 keV) for either of the
two brightness levels that were sampled with the PCA. The seed photon
spectrum is consistent with the emission from a standard disk
(MCD). No direct disk component is required from the fit. In this
interpretation, most of the radiation energy is produced in the
Comptonization corona, which completely covers the inner part of the
disk. Our results basically support the conclusion by \citet{don04},
who analyzed 7 representative PCA spectra of \grs\ in various states
and found that thermal Comptonization is a dominant spectral component
in most (5 out of 7) cases.  In their color-color diagram (see their
Figure~4), our spectra are located at the positions of 
$(1.23, 0.45)$ and $(1.45, 0.52)$, 
which are closest to obs.\ 7 and 6 for Epochs~I' and II',
respectively. \citet{don04} favor a
MCD-dominant model for both obs.\ 7 and 6 
rather than a purely Comptonized spectrum,
however. The reason for this discrepancy is
unclear\footnote{\citet{don04} introduce a strong reflection component
($\Omega_{\rm refl}/2\pi > 2$) in obs.\ 7 to explain the prominent
edge feature around 9 keV as seen in our Epoch~I' spectrum. We infer
that it may be significantly affected by deep absorption K-edges from
highly ionized Fe ions in the disk wind, which are not included in
their spectral model.}.

In the Comptonization model, we can infer the innermost
radius ($r_{\rm in}$) of the accretion disk injecting 
the seed photons to the corona, 
by the conservation of photon number in the Comptonization process. 
For this purpose, we adopt the formula (A1) of \citet{kub04a} 
with the left-hand term
increased by a factor of two,
$$
2\; F_{\rm thc}^{p} \; 2 {\rm cos} (i) = 0.0165\; [\frac{r_{\rm in}^2\;  {\rm cos} (i)}{(D/10\; {\rm kpc})^2}]\; (\frac{T_{\rm in}}{1\; {\rm keV}})^3\;\; {\rm photons\;s^{-1}\;cm^{-2}},
$$ where $F_{\rm thc}^{p}$ is 0.01--100 keV photon flux from the
Comptonized component. Here we have assumed that half of photons in
the corona is injected again into the accretion disk due to the large
optical depth ($\approx 5$). The results are given in Table~4. For
isotropic emission of the Comptonized component, we find $r_{\rm in}$
$\approx$110 km and $\approx$80 km for Epoch~I' and II', respectively
(instead, if the optically thick corona has a thin disk-like geometry,
the $2 {\rm cos} (i)$ factor in the left term must be ignored and
hence $r_{\rm in}$ is increased by about 20\%). The estimated $r_{\rm
in}$ values correspond to $\approx 5 r_{\rm g}$ (Epoch~I') and
$\approx 4 r_{\rm g}$ (Epoch~II'). With correction for the
color-to-effective temperature and that for the inner boundary
condition \citep[see][]{kub98}, the physical radius would be larger
than these values by a factor of 1.2. Thus, based on the innermost
temperature and photon numbers, the optically-thick disk may extend
down only to $\simgt 4 r_{\rm g}$ in this state, not to the innermost
stable circular orbit (ISCO) of $1.23 r_{\rm g}$ for a maximally
rotating Kerr hole. 
If the radius observed in Epoch~II' indeed corresponds to the ISCO,
then our result indicates that the black hole spin of \grs\ is only
moderate, supporting the conclusion by \citet{mid06}.

Of all the classes of light curves for \grs, the $\phi$ class
signifies conditions when the source is most soft and steady. In most
BHBs, these conditions are associated with the high/soft state, in
which a MCD component dominates the X-ray spectrum while there are low
levels of Comptonization \citep{rem06}, normally observed at lower
fractions of the Eddington rate than in our case (0.3--1.2 $L/L_{\rm
Edd}$ depending on the geometry of the emitting region). 
Our analyses show that this is 
not the case for our $\phi$ state observations of \grs , where
Comptonization dominates. Furthermore, the 67 Hz X-ray QPO can be
considered as additional signs of disturbances that lead away from the
simple MCD that we envision for the high/soft state. Such disturbances
are chronic properties of \grs\ in its soft states. \citet{mcc06} were
able to find only 20 high/soft state examples in an archive of $\sim
1000$ RXTE pointings that included $\sim 150$ observations of soft and
steady conditions (classes $\phi$, $\delta$, and some $\gamma$ types).

Such strongly Comptonized spectra are characteristics of the very high
state (VHS) of other BHBs observed at luminosities of $\sim 0.5
L/L_{\rm Edd}$ \citep[e.g.,][]{kub04a}, although the electron
temperature of the Comptonizing corona in the VHS is much higher ($>10
$ keV) than our case ($\approx$4 keV). This suggests that the ``soft
state'' of \grs\ corresponds to the VHS of canonical BHBs, but the
electrons in the corona must be more efficiently cooled by
Comptonization. It could be explained because the disk extends to
small radii in the ``soft state'' of \grs, hence providing many soft
photons to the corona, while the disk of other BHBs in the VHS may be
truncated at larger radii (\citealt{kub04b}; see also
\citealt{don06}). To confirm this picture, more systematic studies of
\grs\ in similar states at different luminosities are important.

There is a puzzling combination of a fast QPO and spectral indications
of strong Comptonization, and this has been seen in other sources as
well \citep{rem06}. If the oscillations originate in the disk, then
the Comptonizing corona must be sufficiently compact to avoid
suppression of the 67 Hz oscillations due to scattering effects; for
$\tau_{\rm e}$=5 the size must be much smaller than $\sim 10 r_{\rm
g}$. Alternatively, the formation of the corona may involve
asymmetries or waves related to energy injection mechanisms. There is
insufficient data to determine if all of the soft-state properties of
\grs\ that deviate from the high/soft state appear to be
correlated. However, such knowledge would assist efforts to understand
the cause of Comptonization and the mechanisms and interpretive values
of the QPO frequencies.


\acknowledgments

We wish to thank Chris Done, Shin Mineshige, and Aya Kubota for useful
discussion. We also thank Herman Marshall for his invaluable advise on
the wavelength calibration of the HETGS, and Jean Swank for
coordinating the simultaneous \rxte\ ToO observations. Part of this
work was financially supported by Grants-in-Aid for Scientific
Research 20540230, and by the grant-in-aid for the Global COE Program
``The Next Generation of Physics, Spun from Universality and
Emergence'' from from the Ministry of Education, Culture, Sports,
Science and Technology (MEXT) of Japan.

\end{document}